\ifx\stupidformat\undefined
\documentclass[12pt]{article}
\else
\documentclass[12pt,a4paper]{artikel1}
\newcommand\auname{\markright{G. J. van Oldenborgh}}
\usepackage{doublespace}
\fi
\def\coloureps{c}
\usepackage{epsfig}
\usepackage[round]{natbib}
\newcommand\dg{${}^\circ$}
\newcommand{\nop}[1]{}

\newlength{\figwidth}
\ifx\stupidformat\undefined\else
\auname
\begin{spacing}{1.66}
\fi
\begin{document}
%
%
\title{What caused the onset of the 1997--1998 El Ni\~{n}o?}
\author{Geert Jan van Oldenborgh\\
\textit{KNMI, De Bilt, The Netherlands}}
\maketitle
\ifx\stupidformat\undefined\else
\mbox{}
\vspace*{\fill}
\begin{flushleft}
Geert Jan van Oldenborgh,\\
KNMI, KS/OO,\\
P.O. Box 201,\\
NL-3720 AE De Bilt,\\
Netherlands
\begin{tabbing}
e-mail: \= oldenbor@knmi.nl\\
tel: \> +31 30 2206711\\
fax: \> +31 30 2202570\\
\end{tabbing}
\end{flushleft}
\clearpage\fi
\begin{abstract}
There has been intense debate about the causes of the 1997--1998 El
Ni\~no.  One side sees the obvious intense westerly wind events as the
main cause for the exceptional heating in summer 1997, the other
emphasizes slower oceanic processes.  We present a quantitative
analysis of all factors contributing to the onset of this El Ni\~no.
At six months' lead time the initial state contributes about 40\% of
the heating compared with an average year, and the wind about 50\%.
Compared with 1996, these contributions are 30\% and 90\%
respectively.  As westerly wind events are difficult to predict, this
limited the predictability of the onset of this El Ni\~no.
\end{abstract}


\section{The Problem}

The 1997--1998 El Ni\~{n}o was one of the strongest on record.
Unfortunately, its onset was not predicted as well as had been hoped
\citep{NSSneakyElNino}.  In spite of claims that an El Ni\~{n}o could be
predicted a year in advance, most predictions 
\citep{AndersonNature,NCEPNinoModel,COLACoupled,KleemanSubsurface}
only started to indicate a weak event six months ahead of time.  
There have therefore been suggestions that El Ni\~no depends not only
on internal factors, but also on external noise in the form of 
weather events in the western Pacific.


The classical picture of El Ni\~{n}o \citep{BjerknesENSO,PhilanderBook}
is that the usual temperature difference between the warm water near
Indonesia and the `cold tongue' in the eastern equatorial Pacific
causes an intensification of the trade winds.  These keep the eastern
region cool by drawing cold water to the surface.  This positive
feedback loop is kept in check by nonlinear effects.  During an El
Ni\~{n}o the loop is broken: a decreased temperature difference causes a
slackening or reversal of the trade winds over large parts of the
Pacific.  This prevents cold water from reaching the surface, keeping
the surface waters warm and sustaining the El Ni\~{n}o.

This picture leaves open the question how an El Ni\~{n}o event is
triggered and terminated.  A variety of mechanisms has been proposed.
On long time scales an unstable mode of the nonlinear coupled
ocean-atmosphere system may be responsible \citep{NeelinSlow}, either
oscillatory or chaotic.  Other authors stress the importance of a
`recharge' mechanism \citep{Wyrtki75,JinRecharge}, with a built-up of
warm water in the western Pacific preceding an El Ni\~no.  Another
description on shorter time scales is in terms of reflections of
equatorial Rossby and Kelvin waves in the thermocline (the interface
between warm surface water and the cold water below at about 100 m
depth).  These would provide the negative feedback that sustains
oscillations 
\citep{SuarezSchopfDelayed,BattistiHirstDelayed,Kessler9193}.
However, short-scale atmospheric `noise' in the form of westerly wind
events in the western Pacific may also be essential in triggering an
El Ni\~{n}o \citep{Wyrtki85,KesslerForcing}.


Here we trace the causes of the onset of last year's El Ni\~{n}o in
May 1997 over the six months from 1 December 1996.  This is the time
scale over which predictions are currently skillful.  Although El
Ni\~no is an oscillation of the coupled ocean-atmosphere system, the
analysis can be simplified by first studying the response of the ocean
to forcing with observed wind stress and heat flux fields.  This
response contains all time delays.  The other part of the loop, the
dependence of the wind stress and heat flux on the ocean surface
temperature will be discussed separately.  

The ocean model used is the Hamburg Ocean Primitive
Equation Model, \textsc{hope} \citep{Frey97,HOPE97} version
2.3, which is very similar to the ocean component of the European
Centre for Medium-range Weather Forecasts (\textsc{ecmwf}) seasonal
prediction system \citep{AndersonNature}, but restricted to the
Pacific Ocean.  It is a general circulation model with a horizontal
resolution of 2.8\dg, increased to 0.5\dg\ along the equator, and a
vertical resolution of 25 m in the upper ocean.  It traces the
evolution of temperature $T$, salinity $S$, horizontal velocities
$u,v$ and sea level $\zeta$.

This ocean model is forced with daily wind stress $(\tau_x,\tau_y)$
and heat flux $Q$ from the \textsc{ecmwf} analysis, which in turn uses
the excellent system of buoys \citep{TOGA1998} that observed this El
Ni\~{n}o.  Evaporation and precipitation are only implemented as a
relaxation to climatological surface salinity.  The initial state
conditions are \textsc{ecmwf} analysed ocean states.  To suppress
systematic model errors we subtract a run starting from an average
1 December ocean state forced with average wind and heat fluxes (both
1979--1996 averages \citep{era1}).

\setlength{\unitlength}{0.1bp}
\ifx\stupidformat\undefined
\begin{figure}[tbp]
\else
\begin{figure}[b]
\fi
\begin{center}
\begin{picture}(3500,1800)(100,250)
\put(0,0){\psfig{file=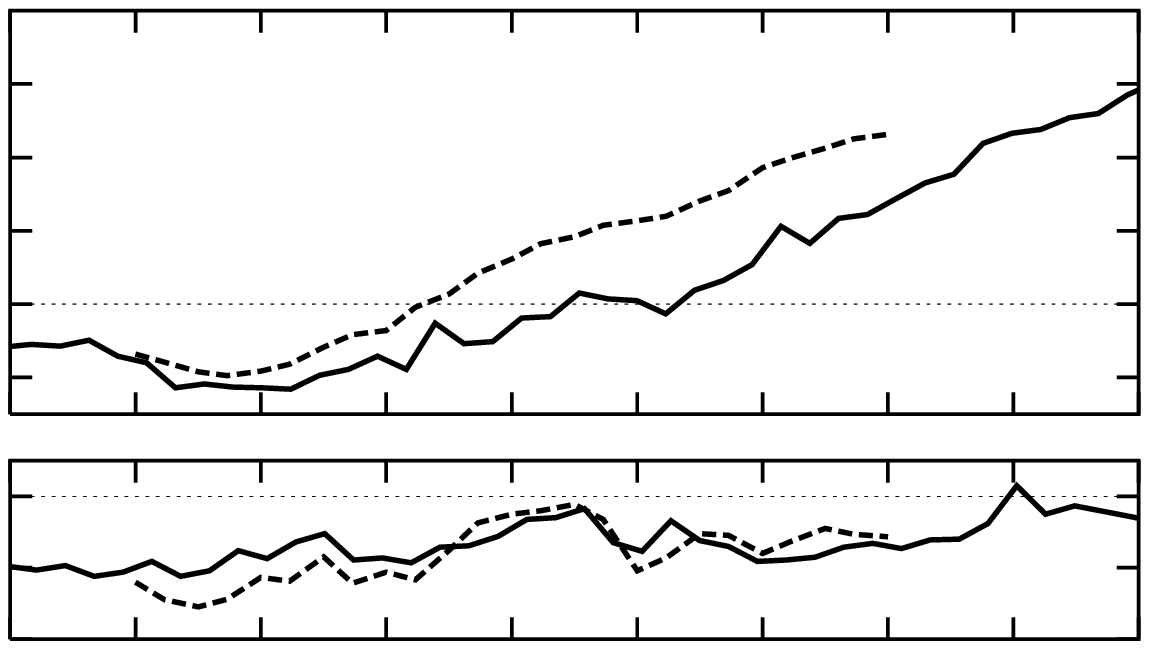}}
\put(650,1949){\makebox(0,0){1996--1997}}
\put(650,580){\makebox(0,0){1995--1996}}
\put(3370,150){\makebox(0,0){J}}
\put(3008,150){\makebox(0,0){J}}
\put(2647,150){\makebox(0,0){M}}
\put(2286,150){\makebox(0,0){A}}
\put(1925,150){\makebox(0,0){M}}
\put(1564,150){\makebox(0,0){F}}
\put(1203,150){\makebox(0,0){J}}
\put(842,150){\makebox(0,0){D}}
\put(481,150){\makebox(0,0){N}}
\put(250,661){\makebox(0,0)[r]{$\mathrm{0}$}}
\put(250,456){\makebox(0,0)[r]{$\mathrm{-1}$}}
\put(250,250){\makebox(0,0)[r]{$\mathrm{-2}$}}
\put(000,2150){\makebox(0,0)[l]{\textsc{nino3} [K]}}
\put(250,1849){\makebox(0,0)[r]{$\mathrm{3}$}}
\put(250,1637){\makebox(0,0)[r]{$\mathrm{2}$}}
\put(250,1426){\makebox(0,0)[r]{$\mathrm{1}$}}
\put(250,1215){\makebox(0,0)[r]{$\mathrm{0}$}}
\put(250,1004){\makebox(0,0)[r]{$\mathrm{-1}$}}
\end{picture}
\end{center}
\caption{The \textsc{nino3} index observed (solid line) and simulated by
the six-month forced model runs (dashed lines).}
\label{fig:ENSOonset}
\end{figure}

The model simulates the onset of the 1997--1998 El
Ni\~{n}o quite well.  We use the \textsc{nino3} index $N_3$, which is a common
measure of the strength of El Ni\~{n}o (the anomalous sea surface
temperature in the area 5\dg S--5\dg N, 90\dg W--150\dg W{}).  In
Fig.~\ref{fig:ENSOonset} the weekly observed \textsc{nino3} index 
\citep{ReynoldsAnalyses} is shown together with the index in the model run, 
compared to the same period one year earlier.  The model
overreacts somewhat to the forcing and simulates a \textsc{nino3}
index of 2.3~K at 1 June 1997, whereas in reality the index reached 
this value one month later.  In 1995--1996 the simulation follows reality very
well.  




\section{The Adjoint Model}

The value of the \textsc{nino3} index at the end of a model run can be
traced back to the model input (initial state, forcing) with an
\emph{adjoint model}.  The normal ocean model is a (complicated)
function $\mathcal{M}$ that takes as input the state of the ocean at
some time $t_0$ (temperature $T_0$, salinity $S_0$, etc.).  Using the
wind stress $\vec{\tau}_i$ and heat flux $Q_i$ for each day $i$ for
six months it then produces a final state temperature $T_n$.  The
adjoint model (or backward derivative model) is
the related function that takes as input derivatives to a scalar function 
of the final state, here the \textsc{nino3} index, $\partial
N_3/\partial T_n$.  It goes backward in time and uses the chain rule of 
differentiation \citep{GieringRecipes} to compute from these
(and the forward trajectory) the derivatives 
$\partial N_3/\partial T_0$, $\partial
N_3/\partial S_0$, $\partial N_3/\partial\vec{\tau}_i$ and $\partial
N_3/\partial Q_i$.   These derivatives can be interpreted as
\emph{sensitivity fields},  
giving the effect of a perturbation in the initial state or forcing fields.
We can use them to make a Taylor expansion of the
\textsc{nino3} index to all the input variables:
\begin{eqnarray}
\label{eq:dN3}
N_3 & \approx & \frac{\partial N_3}{\partial T_0}\cdot\delta T_0 
        + \frac{\partial N_3}{\partial S_0}\cdot\delta S_0 
\nonumber\\ && \mbox{}
        + \sum_{\mathrm{days}~i} \left(
        \frac{\partial N_3}{\partial\vec{\tau}_i}\cdot\delta\vec{\tau}_i
      + \frac{\partial N_3}{\partial Q_i}\cdot\delta Q_i
                \right)
\end{eqnarray}
This means that the value of the index is explained as a sum of the
influences of initial state temperature and salinity, and the wind and
heat forcing during the six months of the run.  These influences are
each a dot product of the sensitivity to this variable (computed with
the adjoint model) multiplied by its deviation from the normal state
(extracted from the \textsc{ecmwf} analyses).  To minimize higher
order terms we take the average derivative from the simulation and the
climatology run.  We have checked with actual perturbations that the
accuracy of the linear approximation Eq.~\ref{eq:dN3} is usually 
better than about 30\% (within the model).  Details can be found in
\citet{OldenborghTracking}.


\section{The 1997--1998 El Ni\~{n}o}

For the value of the \textsc{nino3} index on 1 June 1997 the
linearization Eq.~\ref{eq:dN3} gives a value of 1.8~K, compared with
the 2.3~K simulated (and 1.3~K observed), this is within the expected
error.  The high value is mainly due to the influence of the westerly
wind anomalies (1.0~K) and the initial state temperature on 1 December
1996 (1.1~K){}.  The salinity contributes $-$0.3~K, with a large
uncertainty.

\begin{figure}
\begin{center}
\figwidth=0.4\textwidth
\makebox[0cm][r]{\Large a}\raisebox{3ex}%
{\psfig{file=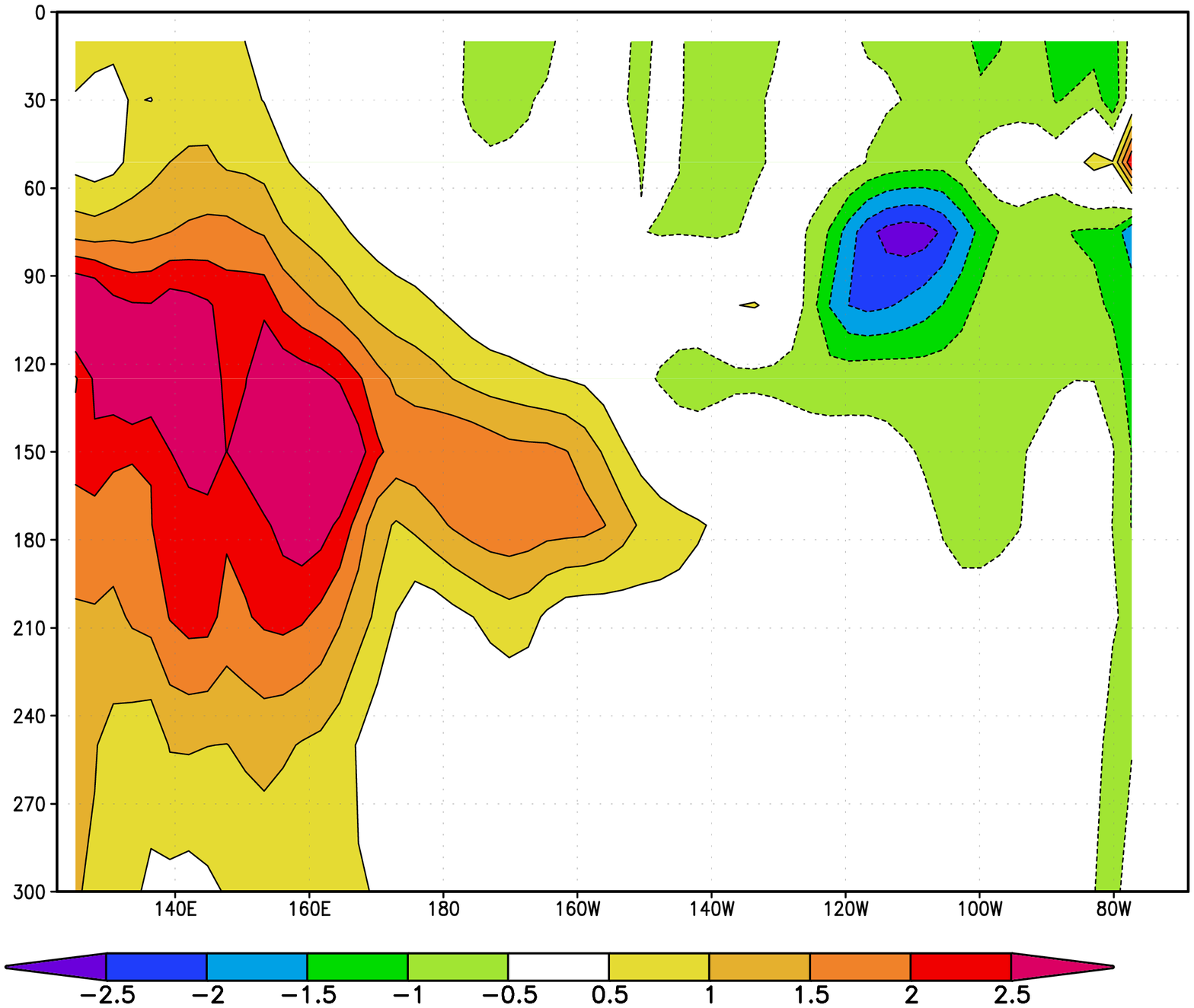,angle=-90,width=\figwidth}}\\[2mm]
\hspace*{\figwidth}\makebox[0cm][r]{$\delta T_0\:[\mathrm{K}]$}\\
\makebox[0cm][r]{\Large b}\raisebox{3ex}%
{\psfig{file=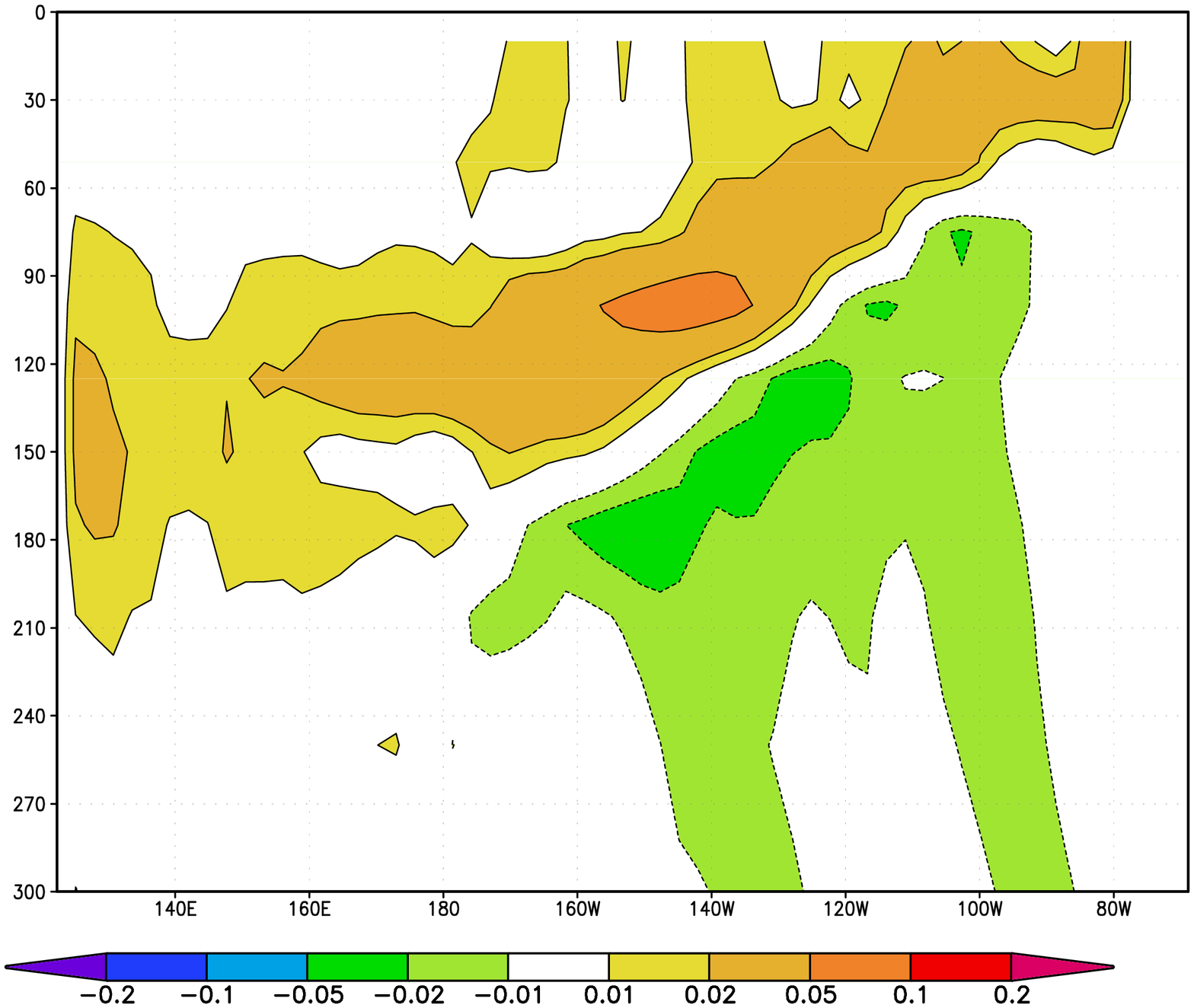,angle=-90,width=\figwidth}}\\[2mm]
\hspace*{\figwidth}\makebox[0cm][r]{$\partial N_3/\partial T_0\:[\mathrm{K/K/sr/m}]$}\\
\makebox[0cm][r]{\Large c}\raisebox{3ex}%
{\psfig{file=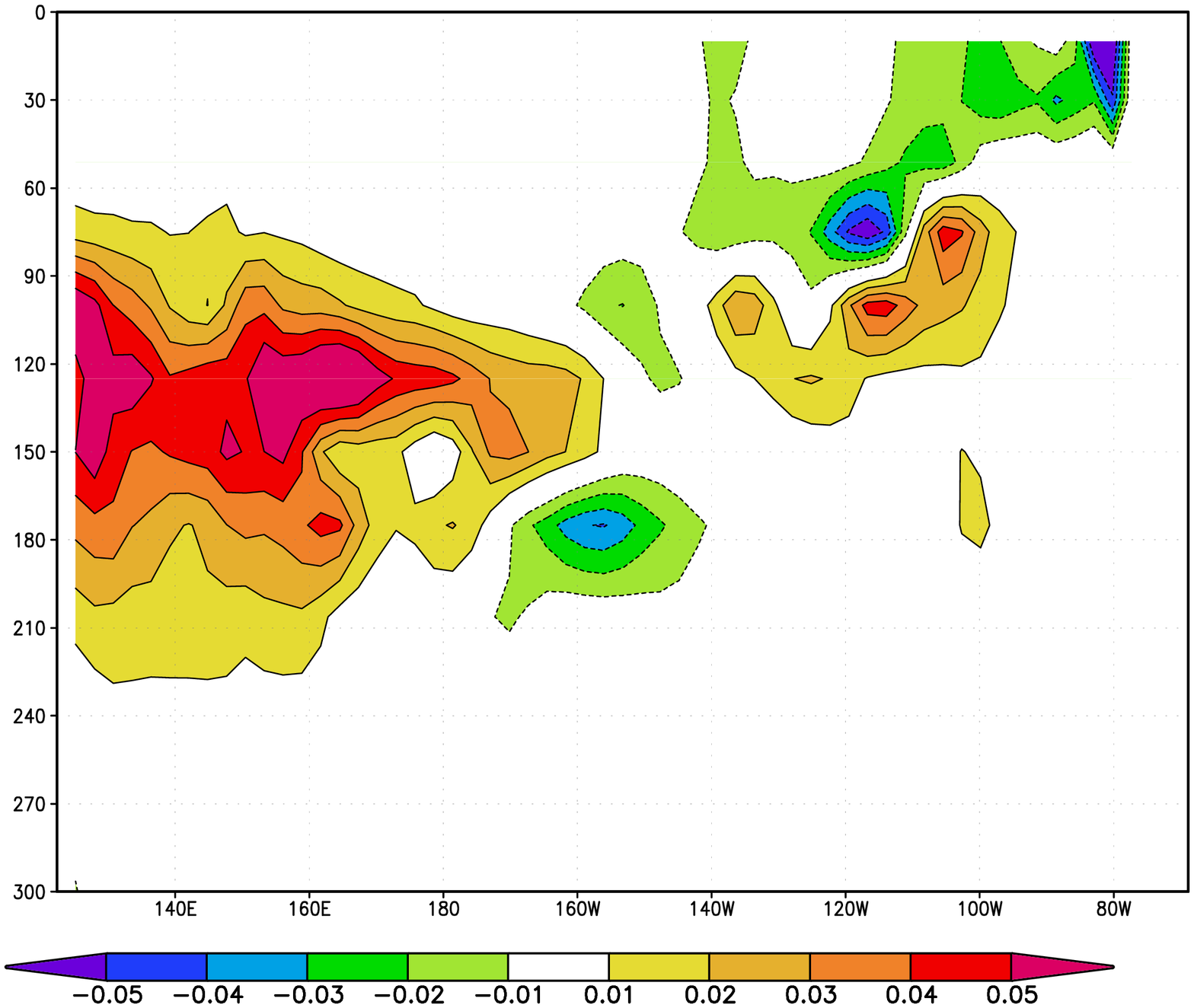,angle=-90,width=\figwidth}}\\[2mm]
\hspace*{\figwidth}\makebox[0cm][r]{$\partial N_3/\partial T_0\cdot\delta T_0\:[\mathrm{K/sr/m}]$}
\end{center}
\caption{Depth-longitude plots of the effect of the initial state
temperature on the \textsc{nino3} index in early June.  At the top the
analyzed temperature anomalies (averaged over 5\dg S--5\dg N) are
shown at the beginning of December 1996; the second frame depicts the
sensitivity of the ocean to these temperature anomalies and the third
the product of these two, which gives the rise in the \textsc{nino3}
index on June 1 due to the thermal structure six months earlier.}
\label{fig:Tini}
\end{figure}

The spatial structure of the influence of the initial state
temperature is shown in Fig.~\ref{fig:Tini}.  The top panel gives the
temperature anomaly $\delta T_0$ along the equator at the
beginning of the run (Dec 1996), showing an unusually deep
thermocline in the western Pacific and a shallower thermocline in the
eastern Pacific.  The second frame depicts the sensitivity of the June
\textsc{nino3} index to temperature anomalies six months earlier,
$\partial N_3/\partial T_0$.  The third frame is just the product of
the previous two; the integral of this over the whole ocean gives the
1.1~K contribution to the \textsc{nino3} index mentioned before.  The
contribution is concentrated in the deeper layer of warm water along
the equator in the western Pacific, in agreement with a `recharge'
mechanism.

\begin{figure}[tb]
\begin{center}
\begin{picture}(3800,1250)(100,750)
\put(0,0){\psfig{file=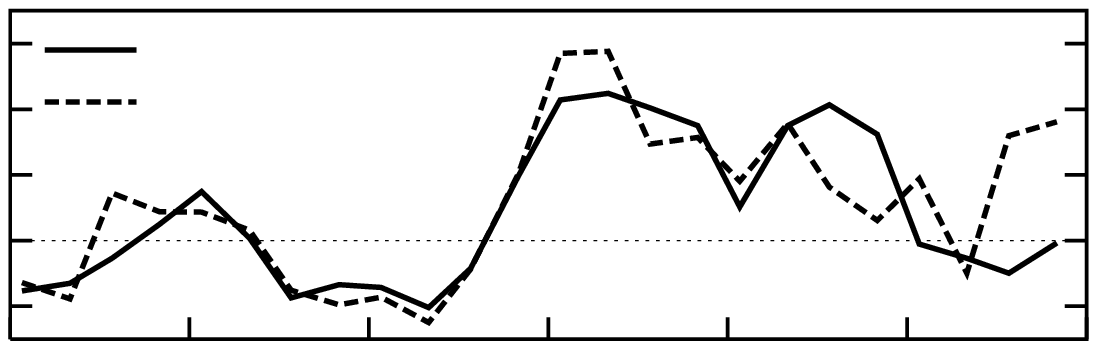}}
\put(3291,800){\makebox(0,0){M}}
\put(2775,800){\makebox(0,0){A}}
\put(2258,800){\makebox(0,0){M}}
\put(1741,800){\makebox(0,0){F}}
\put(1225,800){\makebox(0,0){J}}
\put(1225,650){\makebox(0,0){1997}}
\put(708,800){\makebox(0,0){D}}
\put(708,650){\makebox(0,0){1996}}
\put(863,1731){\makebox(0,0)[l]{$\partial N_3/\partial\tau_x\cdot\delta\tau_x$}}
\put(100,1950){\makebox(0,0)[l]{[K/week]}}
\put(400,1749){\makebox(0,0)[r]{$\mathrm{0.3}$}}
\put(400,1560){\makebox(0,0)[r]{$\mathrm{0.2}$}}
\put(400,1371){\makebox(0,0)[r]{$\mathrm{0.1}$}}
\put(400,1182){\makebox(0,0)[r]{$\mathrm{0}$}}
\put(400,993){\makebox(0,0)[r]{$\mathrm{-0.1}$}}
\put(863,1581){\makebox(0,0)[l]{$\delta\tau_x$}}
\put(3900,1950){\makebox(0,0)[r]{[$\mathrm{Nm^{-2}}$]}}
\put(3900,1749){\makebox(0,0)[r]{$\mathrm{0.06}$}}
\put(3900,1560){\makebox(0,0)[r]{$\mathrm{0.04}$}}
\put(3900,1371){\makebox(0,0)[r]{$\mathrm{0.02}$}}
\put(3900,1182){\makebox(0,0)[r]{$\mathrm{0}$}}
\put(3900,993){\makebox(0,0)[r]{$\mathrm{-0.02}$}}
\end{picture}
\end{center}
\caption{The influence of the zonal wind stress $\tau_x$ on the
\textsc{nino3} index at 1 June 1997 during the previous six months
(solid line), the average anomalous wind stress over the area 130\dg E to 
160\dg W, 5\dg S to 5\dg N.}
\label{fig:influences_onset}
\end{figure}

Fig.~\ref{fig:influences_onset} shows the time structure of the
influence of the zonal wind stress.  The area under the solid graph
gives the total influence, 1.0~K.  The main causes of warming are the
three peaks in zonal wind stress (dashed line) at the beginning of
March, the end of March and the beginning of April, contributing about
0.6~K, 0.3~K and 0.5~K respectively.  The peaks correspond with (very)
strong westerly wind events in the western Pacific.  These generated
downwelling Kelvin waves in the thermocline that travelled east and
deepened the layer of warm water in the eastern Pacific 2--3 months
later, increasing the surface temperature.  There was also a strong
wind event in December, contributing about 0.4~K over a negative
baseline.  From Fig.~\ref{fig:influences_onset} it seems likely that
it increased the strength of the later wind events by heating the
eastern Pacific in March.  The heating effect of the March wind event
also gave rise to an increase of the wind stress $\delta\tau_x$ in
May, but this reversal of the trade winds does not yet influence the
\textsc{nino3} index $\partial N_3/\partial\tau_x\cdot\delta\tau_x$,
justifying the uncoupled analysis.

\begin{figure}
\begin{center}
\figwidth=0.8\textwidth
\makebox[0cm][r]{\Large a}\raisebox{3ex}%
{\psfig{file=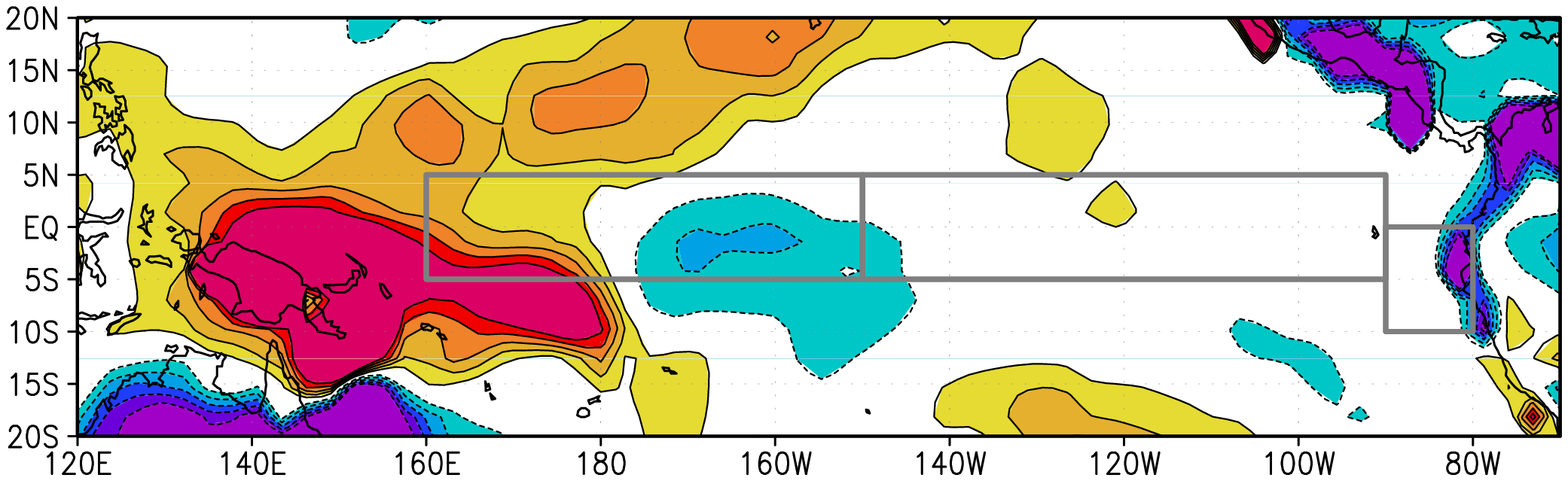,angle=-90,width=\figwidth}}\\[-4mm]
\psfig{file=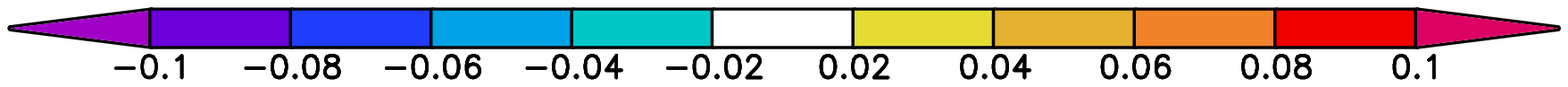,angle=-90,width=\figwidth}\\[0mm]
\hspace*{\figwidth}\makebox[0cm][r]{$\delta\tau_x\:[\mathrm{Nm^{-2}}]$}\\[-2mm]
\makebox[0cm][r]{\Large b}\raisebox{3ex}%
{\psfig{file=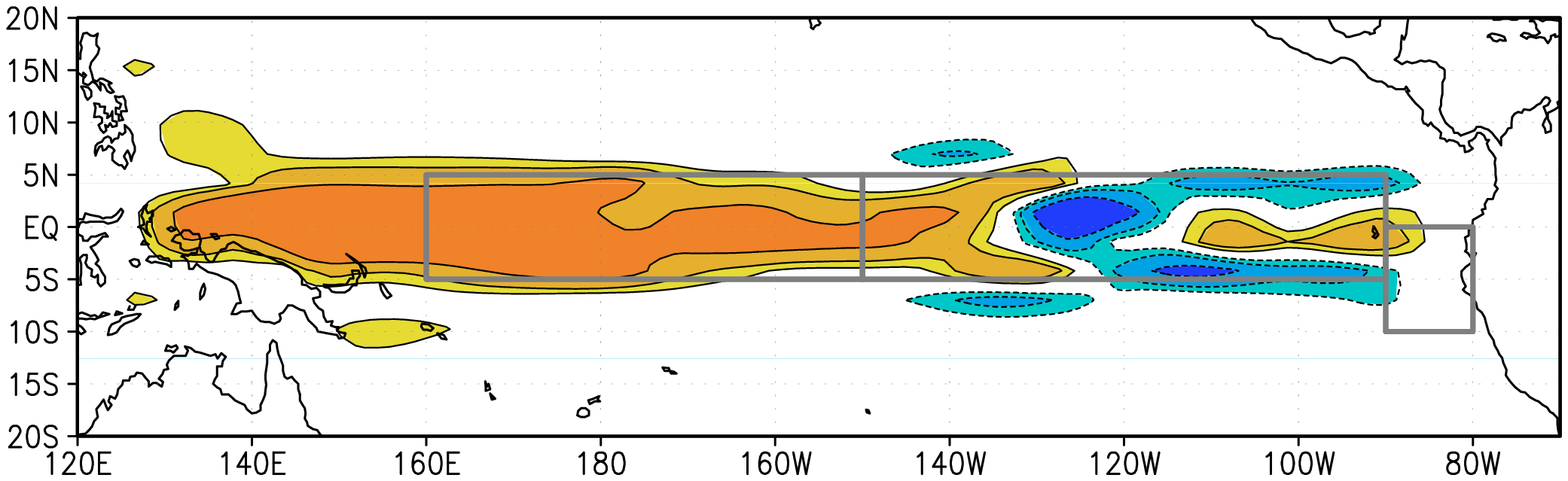,angle=-90,width=\figwidth}}\\[-4mm]
\psfig{file=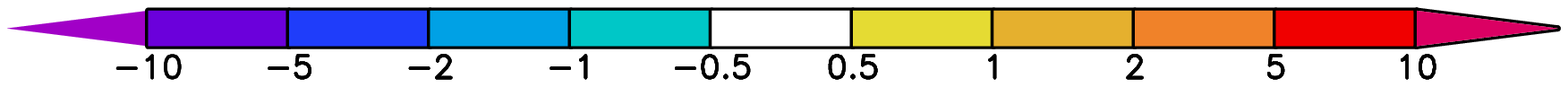,angle=-90,width=\figwidth}\\[0mm]
\hspace*{\figwidth}\makebox[0cm][r]{$\partial N_3/\partial \tau_x\:[\mathrm{K/Nm^{-2}/sr/day}]$}\\[-2mm]
\makebox[0cm][r]{\Large c}\raisebox{3ex}%
{\psfig{file=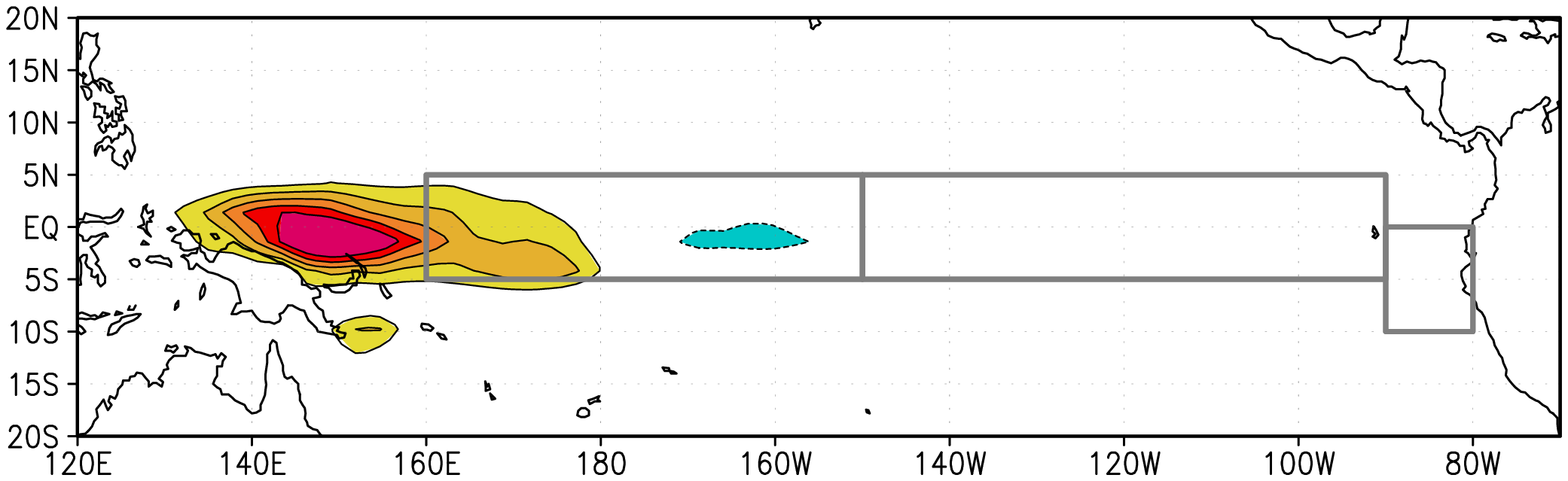,angle=-90,width=\figwidth}}\\[-4mm]
\psfig{file=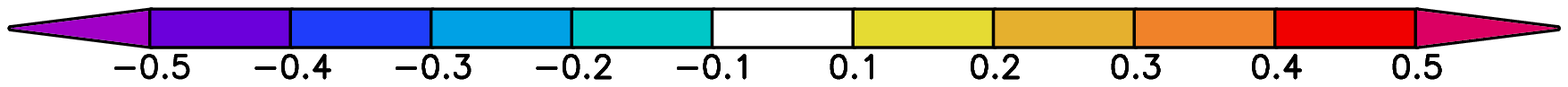,angle=-90,width=\figwidth}\\[0mm]
\hspace*{\figwidth}\makebox[0cm][r]{$\partial N_3/\partial \tau_x\cdot\delta\tau_x\:[\mathrm{K/sr/day}]$}\\[-2mm]
\end{center}
\caption{The effect of the March westerly windburst on the
\textsc{nino3} index in early June.  At the top the averaged westerly
wind stress anomaly for the week centered on 11 March 1997 is shown,
the second frame depicts the sensitivity of the ocean to zonal
wind stress and the third the product of these two which gives the
rise in the \textsc{nino3} index on June 1 due to this wind event.}
\label{fig:march}
\end{figure}

The structure of the peaks in Fig.~\ref{fig:influences_onset} can be
seen more clearly in spatial views.  In Fig.~\ref{fig:march}a the
zonal wind stress anomaly $\delta\tau_x$ is plotted for the second
week of March.  The westerly wind event corresponds to the large
localized westerly anomaly around 150\dg E{}.  Fig.~\ref{fig:march}b
shows the sensitivity of the \textsc{nino3} index in June to the zonal
wind stress during this week, $\partial N_3/\partial\tau_x$.  This
sensitivity consists of two main parts, both equatorally confined.  In
the western and central Pacific extra westerly wind stress would
excite a downwelling Kelvin wave, raising the \textsc{nino3} index
three months later.  In the eastern Pacific the response would be in
the form of a Rossby wave.  The product of the anomaly and sensitivity
fields is shown in Fig.~\ref{fig:march}c.  This gives the influence of
zonal wind stress during this week on the \textsc{nino3} index, the
integral of this field gives the corresponding value (0.22~K) in
Fig.~\ref{fig:influences_onset}.  The influence is contained in the
intersection of the westerly wind event and the equatorial wave guide,
and very localized in time and space.  The effects of the other wind
events are similar.

The question remains whether the big influence of these wind events
was due to their strength $\delta\tau_x$ or to an increased
sensitivity of the ocean $\partial N_3/\partial\tau_x$.  We therefore
repeated the analysis for the same months one year earlier, when the
temperature in the eastern Pacific stayed below normal
(Fig.~\ref{fig:ENSOonset}).  The adjoint model gives a \textsc{nino3}
index of $\mathrm{-0.6}$~K, equal to the simulated index (the observed
index was $\mathrm{-0.7}$~K).  This index is built up by a large
negative influence of the wind stress, $\mathrm{-1.5}$~K, and a
positive influence of the heat flux, $\mathrm{+0.9}$~K.  The influence
of the initial state temperature is also positive, but weaker than in
the 1996--1997 $\mathrm{+0.6}$~K, and the salinity contributes
$\mathrm{-0.5}$~K.

Although the built-up of warm water is also less pronounced, 
the largest difference is in the influence of the zonal
wind stress.  The sensitivity to zonal wind stress $\partial
N_3/\partial\tau_x$ (over the area where its variability is largest)
is compared for these two years in Fig.~\ref{fig:avedflux}.  During
the time of the strong early March windevent the sensitivity was not
very different bewteen the two years, but it was a factor two higher
in April 1997 than in April 1996, and lower during the first two
months.  In all, these differences cannot explain more than a few
tenths of a degree difference in the \textsc{nino3} index on 1 June.

\begin{figure}[tb]
\begin{center}
\begin{picture}(3600,1250)(0,200)
\put(0,0){\psfig{file=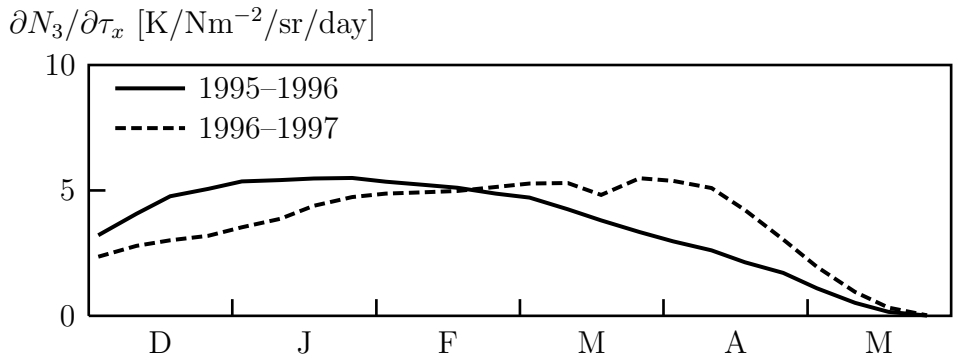}}
\put(713,958){\makebox(0,0)[l]{1996--1997}}
\put(713,1108){\makebox(0,0)[l]{1995--1996}}
\put(3279,150){\makebox(0,0){M}}
\put(2738,150){\makebox(0,0){A}}
\put(2196,150){\makebox(0,0){M}}
\put(1654,150){\makebox(0,0){F}}
\put(1113,150){\makebox(0,0){J}}
\put(571,150){\makebox(0,0){D}}
\put(000,1350){\makebox(0,0)[l]{$\partial N_3/\partial\tau_x$
[$\mathrm{K/Nm^{-2}/sr/day}$]}}
\put(250,1196){\makebox(0,0)[r]{$\mathrm{10}$}}
\put(250,723){\makebox(0,0)[r]{$\mathrm{5}$}}
\put(250,250){\makebox(0,0)[r]{$\mathrm{0}$}}
\end{picture}
\end{center}
\caption{The average sensitivity of the \textsc{nino3} index on 1 June
to westerly winds in the area defined in
Fig.~\ref{fig:influences_onset}.}
\label{fig:avedflux}
\end{figure}

The difference between an El Ni\~{n}o in 1997 and no El Ni\~no in 1996
can be attributed for about 30\% to an even stronger built-up of warm
water in the western Pacific, and for about 90\% to the the absence of
strong westerly wind events in the western Pacific in the 1995--1996
rain season.  A successful prediction scheme will have to predict the
intensity of the westerly wind events correctly.  However, the
year-to-year variability of these wind events does not seem to depend
on the state of the Pacific ocean \citep{Slingo98}, and at the moment
is not predictable.


\section{Conclusions}

Using an adjoint ocean model we have shown that a successful
prediction of the strong onset of the 1997--1998 El Ni\~{n}o, required
a successful prediction of strong westerly wind events in
March--April, which in our model contributed about 90\% to the strength of
the El Ni\~no on 1 June 1997 compared to the situation one year earlier.
The sensitivity to these wind events was not significantly different
from the year before.  The built-up of warm water contributed about
30\% of the difference.  The strong dependence on the westerly wind
events would explain the relatively short lead time for correct
predictions of the strong onset of this El Ni\~{n}o.


\paragraph{Acknowledgments} I would like to thank the \textsc{ecmwf} seasonal
prediction group for their help and support and Gerrit Burgers for his
part in the construction of the adjoint model.  This research was
supported by the Netherlands Organization for Scientific Research
(NWO).


\ifx\stupidformat\undefined\else
\newpage
\listoffigures
\end{spacing}
\fi
\end{document}